\documentclass[12pt,preprint]{aastex}

\bibpunct{(}{)}{;}{a}{}{,} 

\shorttitle{The discovery of the Rosette Eye} \shortauthors{Li et
al.}

\begin{document}

\title{The Rosette Eye: the key transition phase in the birth of a massive star}
\author{J. Z. Li}
\affil{National Astronomical Observatories, Chinese Academy of
Sciences, Beijing 100012, China; ljz@bao.ac.cn}

\author{M. D. Smith}
\affil{Centre for Astrophysics \& Planetary Science, University of
Kent, Canterbury CT2 7NH, UK}

\author{R. Gredel}
\affil{Max-Planck Institut f\"{u}r Astronomie, K\"{o}nigstuhl 17,
D-69117 Heidelberg, Germany}

\author{C. J. Davis}
\affil{Joint Astronomy Centre, 660 North A`ohoku Place, Hilo, HI
96720}

\author{T. A. Rector}
\affil{University of Alaska at Anchorage, 3211 Providence Drive,
Anchorage, AK 99508}

\begin{abstract}

Massive protostars dramatically influence their surroundings via
accretion-induced outflows and intense radiation fields. They
evolve rapidly, the disk and infalling envelope being evaporated
and dissipated in $\sim$~ 10$^5$ years. Consequently, they are
very rare and investigating this important phase of early stellar
evolution is extremely difficult. Here we present the discovery of
a key transient phase in the emergence of a massive young star, in
which ultraviolet radiation from the new-born giant has just
punctured through its natal core. The massive young stellar object
AFGL 961 II is readily resolved in the near infrared. Its
morphology closely resembles a cat's eye and is here dubbed as the
Rosette Eye. Emerging ionized flows blow out an hourglass shaped
nebula, which, along with the existence of strong near-infrared
excess, suggests the existence of an accretion disk in the
perpendicular direction. The lobes of the hourglass, however, are
capped with arcs of static H$_{2}$ emission produced by
fluorescence. This study has strong implications for our
understanding of how massive stars embark on their formation.

\end{abstract}

\section{Introduction}

Protostellar objects are, in most cases, deeply embedded in
molecular clouds and are enshrouded by heavy foreground
extinction. This makes the very early stages of stellar gestation
notoriously illusive~(Zinnecker \& Yorke 2007). The birth of stars
with masses above 10~M$_\odot$ is particularly intriguing as their
radiation is apparently sufficient to resist the accretion of gas and,
hence, further mass growth unless somehow mitigated~(Mckee \& Tan
2002; Behrend \& Maeder 2001). Evidence of potential disks and/or
envelopes associated with massive star formation has been
accumulated through various probes especially in the radio
domain~(Chini et al. 2004; Patel et al. 2005). However, the
formation process of massive stars remains far from being
resolved. The critical growth period of massive stars lasts only
tens of thousand of years but is usually accompanied by spectacular
ejections of gas in opposite directions. Since such jets are
circumstantial evidence of an accretion disk, it is possible and
crucial to gain robust evidence of ongoing accretion associated
with massive young stellar objects (YSO) in their early stages of
evolution.

The Rosette Molecular Complex (RMC) is a famous isolated massive
star forming region with an extent of about 100 pc. It is located
at a distance of $\sim$~1.39 kpc (Hensberge et al. 2000) and
harbors a gas reservoir of $\sim$10$^5~M_\odot$(Blitz \& Thaddeus
1980). New generation OB star formation is in evidence in the
densest ridge of the Complex (Li \& Smith, 2005). The
well known high-mass protostellar system AFGL 961 is situated
well within this region (Cohen 1973; Grasdalen et al. 1983; Castelaz
et al. 1985; Lenzen et al. 1984; Hodapp 1994). The third component
of the system, designated as AFGL 961 II (Li \& Smith 2005), was
first noticed to be associated with a small nebulosity by Eiroa
(1981) and later briefly discussed by Hodapp (1994). The
nebulosity was considered as a cavity with a partial shell
surrounding the central young star~(Aspin 1998; Alvarez et al.
2004). However, the origin of the intriguing YSO is far from
clear.

\section{Observations and Data Reduction}

\subsection{Infrared imaging and spectroscopy}

We obtained near-infrared images of the AFGL 961 region as part of
our program to explore the entire Rosette star formation complex.
The JHK and H$_2$ data were obtained on December 19-20, 2005, with
the SOFI instrument (Moorwood, Cuby \& Lidman 1998) on the ESO New
Technology Telescope at La Silla in Chile. The integration time is
2000s in the H$_2$~1--0\, S(1) and 300s in each of the J, H and Ks
bands. We also employed SOFI to obtain broad Ks-band spectroscopy.
The HR Grism at order 3 results in a medium spectral resolution of
$\sim$\,2200 over the wavelength range 2.00-2.30 $\mu$m. A
0$\arcsec$.6 slit was used, cutting through both the exciting
source and the bipolar shock structures at a position angle of
345$^{\circ}$ (north to east). Four 300s exposures of the target
source were taken, which was accompanied by a 300s exposure of the
blank sky.  A telluric standard star (Hip 1179 with a spectral
type of A7V, Perryman et al. 1997) was observed right after the
exposures of the target. The integration times are 3 x 60\,s for the
standard star and 60\,s for the sky background.


Near-infrared echelle spectra covering H$_2$ 1-0\,S(1) were
obtained at the U.K. Infrared Telescope (UKIRT) on September 18,
2006 UT. The cooled grating spectrometer CGS\,4 (Mountain et al.
1990) was used, which employs a 256$\times$256 pixel InSb array
and has a pixel scale of 0.41"\,$\times$\,0.88" (0.41" in the
dispersion direction).  A 2-pixel-wide slit was used which yields
a velocity resolution of $\sim$16~km\,s$^{-1}$. An internal
black-body lamp was used to flat-field each spectral image, before
the sky frames were subtracted from each object frame. The coadded
spectral images were then wavelength calibrated using sky lines in
the sky frame. The overall velocity calibration is accurate to
better than 6~km\,s$^{-1}$ , while velocity shifts between
adjacent spectra observed along the same slit are accurate to
within $\sim$2~km s$^{-1}$.  Finally, observations of HD\,42807
(BS\,2208; G2V, V\,=\,6.22~mag) obtained immediately before the
data were used to correct for telluric absorption and to flux
calibrate each spectral image.

\subsection{Optical imaging}

The AFGL 961 system was observed on the night of 15 December 2006
with the Kitt Peak National Observatory 4-meter telescope and the
Mosaic I camera (Muller et al. 1998). The pixel scale is
0.258\arcsec pixel$^{-1}$. Five 600\,s exposures were obtained
in each of the H$\alpha$ (k1009) and H$\alpha$+16nm/[SII] (k1013)
filters. The median seeing in both filters is about 0.9\arcsec.
The exposures were dithered so that the gaps between the CCDs were
filled during the stacking process. The data were reduced with the
IRAF MSCRED package in the standard manner.

\subsection{Optical spectroscopy}

Medium resolution spectroscopy of AFGL 961 II was performed on the
night of 22 January 2006 with the 2.16 m telescope of the National
Astronomical Observatory of the Chinese Academy of Sciences
(NAOC). An OMR (Optomechanics Rsearch Inc.) spectrograph and a
Tecktronix 1024 x 1024 CCD were used. A 50~\AA~mm$^{-1}$ grating
and a 2 $\arcsec$ slit resulted in a two-pixel resolution of the
spectra of 2.4~\AA. The accuracy of the wavelength calibration
allows sampling of velocities down to 20 km s$^{-1}$.

\section{Results}

We resolve the nebulous YSO, AFGL 961~II, with our data into distinctive
structures. Its spectacular appearance closely resembles a cat's eye
in our NIR colour-composite image (Fig.~1) and we thus refer to it
as the Rosette Eye. It is composed of a bright young star at the
center, an hourglass shaped diffuse nebula oriented at a position
angle of 345 $^\circ$ (from north to east), and extensive
molecular emission resembling bipolar shock structures on either
side of the apparent central source. The molecular emission is the
most prominent in the well-defined arc-like structures of the
system. A faint star is also found in close proximity to the
central star (see Fig. 1), but does not appear to be physically
related to the excitation of the extended nebulosity. The spatial
appearance of the Eye, along with the detection of strong excesses
in the near-infrared (Li \& Smith, 2005), suggests the existence
of an accretion disk in the perpendicular direction.

At a distance of 1.39 kpc (Hensberge et al. 2000), the arc
structures arch over an hourglass shaped cavity with a physical
scale of $\sim$~0.087 pc, which is in agreement with typical sizes
of $\sim$ 0.1~pc of massive star-forming clumps~(Kurtz, et al.
2000; Motte \& Andre 2001).
If we assume a molecular density of n$_{H_2}$ = 10$^7$ cm$^{-3}$
(De Pree et al. 1998, Motte et al. 2001, Churchwell 2002), a
pre-stellar core of this size will harbor a gas reservoir of over
53.5 M$_\odot$ and is rich enough to give birth to a 20~M$_\odot$
(Li \& Smith 2005) infant star according to theoretical
models~(Yorke \& Sonnhalter 2002).

The spatial appearance of the Rosette Eye in each of the observed
bands is presented in Fig.~2, where we see a clear bipolar nebula
in J, resembling an hourglass with an opening angle of about
60$^\circ$. An eye structure is superimposed; the north-west arc
is more distinctively attached to the end of the conical nebula
than is the south-east arc. There is also diffuse scattered light
marginally visible beyond the north-west arc. The conical nebula
to the south-east possesses a sharp edge, suggesting the existence
of high extinction in that direction. Diffuse extensive emission
extends further to the east, pointing toward the massive binary
associated with AFGL\,961, probably a leakage of the stellar light
over the cavity walls. The hourglass-shaped nebula is less
prominent in the H, Ks and H$_{2}$ bands with the apparent opening
angle reduced to about 45$^\circ$. The arc structure in the
south-east, however, brightens remarkably as the wavelength
increases. It is sharp and bright in the Ks band image and is most
prominent in the H$_{2}$ narrow-band image.

In the optical narrow-band H$\alpha$, the structure to the
north-west resembles that observed in the J band but displays more diffuse
and extended emission. The south-east lobe appears to be dominated
by heavy extinction and is restricted to the Eye structure. Our
mid-resolution optical spectroscopy of the exciting source shows
very strong H$\alpha$ emission (EW=170\AA, FWHM=$\sim$500 km
s$^{-1}$). This, along with the broad Br\,$\gamma$ emission
presented below, is a good indicator of youth as also seen in
other massive YSOs (Bunn et al. 1995). However, only marginal
[SII] emission is detected from both the central YSO and the
photoionized nebula. The spectroscopy with a resolution of
1.0~\AA~pixel$^{-1}$ yields no velocity difference between the
bipolar lobes. This suggests a low velocity of at most a few tens
of km~s$^{-1}$ of the outflowing gas, in accord with the general
properties of outflows associated with massive protostellar
objects (Churchwell 2002; Henning et al., 2000).

Given the optical
spectroscopy results, the [SII] image closely represents continuum
emission from the Eye. The derived continuum-subtracted H$\alpha$
emission shows a small HII region excited by the UV ionizing
photons from the YSO and faint net emission from the north-west
arc, indicating marginal ionization in the shell. The small HII
region is confined by the bipolar cavity, which is less than 0.1
pc in size. Here the UV ionizing photons are attributed to shocked
accretion from the proposed disk/envelope in the orthogonal
direction, which feeds the YSO continuously till the final
dissipation of the circumstellar materials. The morphology of the
Eye matches well the transient phase in massive stellar evolution
presented by Keto 2007, where an hourglass-shaped
photon-dissipation region first forms before it evolves into a
full-blown HII region. The excitation of the ionized nebula,
however, suggests a spectral type earlier than B2 for the exciting
star, commensurate with its position on the color-magnitude
diagram in the NIR (Li \& Smith 2005).


Archived Spitzer IRAC and MIPS imaging data on the AFGL~961
region, which traces hot dust distribution, indicate that
AFGL\,961~II, the exciting source of the Eye, sits on the edge of
a lane of dust with the highest extinction. The
conical structure in the south-east encounters heavy dust
extinction. This explains why the bipolar nebula is observed in the
optical as a fan shaped nebula extending toward the north-west~(Li
\& Smith 2005). The MIPS 24~$\mu$m imaging, on the other hand,
indicates a dumbbell shaped nebula that encompasses the bipolar
cavity detected in the NIR, which is devoid of dust emission and
implies the dissipation or evacuation of dust by the ionized
flows. With a moderate extinction of $\sim$~7~mag (Li \& Smith
2005) and a favorable viewing angle, we propose that AFGL\,961~II
represents a spectacular case of a massive star visible in its
early stage of formation.

Strong H$_{2}$ line emission often arises after collisional
excitation within shock waves in association with mass ejections
from extremely young stars. This usually yields relatively high
radial velocities and strong excitation of the lower vibrational
levels. To test this hypothesis, we performed deep Ks band
spectroscopy with a slit position cutting through both the arc
structures and the central YSO. The spectrum of the exciting
source shows Br\,$\gamma$ (EW=10.8\AA, FWHM=29\AA) superimposed on
a featureless rising continuum towards longer wavelengths.
However, the H$_{2}$ emission from the shell displays no sign of
collisional excitation. Line emission originates from very high
vibrational levels of H$_2$ (Fig.~3, upper panel). The highest
level present is the 9--7~Q(3) at 2.100\,$\mu$m, originating
42,462\,K above the ground state. The spectroscopic data
demonstrate distinctly that the gas is purely fluoresced with the
H$_2$ line strengths corresponding exactly to an energy level
cascade. This, however, is in agreement with theoretical
predictions that UV pumped fluorescence should be a common
phenomenon, especially on surfaces of molecular clumps illuminated
by young massive stars~(Gatley 1987). AFGL\,961~II thus represents
the first detection of pure fluorescent radiation directly
associated with massive protostellar objects.

The fluorescence
origin of the line emission is confirmed by our simulation based
on the observed line flux ratios (Smith, Li \& Gredel et al. in
prep.). Furthermore, the echelle spectrogram obtained by UKIRT
discloses no radial velocity motion in the shell to the limits of
our measurements ($\sim$~6~km\,s$^{-1}$) (Fig.~3, lower panel).

\section{Summary \& discussion}

This study substantiates that the exciting source of the Rosette
Eye, AFGL 961 II, has just finished its original collapse, that UV
radiation from shocked massive accretion has recently turned on
and that ionized stellar winds have begun to emerge in the polar
directions. The shell left by the original collapse is now
prepared to face the UV ionization from the newly born star. We
suggest that the Eye may closely correspond to a transient phase
immediately preceding that of the famous Orion OMC-1
outflow~(Stone et al. 1995). In this scenario, the arcs will be
subjected to fluid instabilities as an ensuing fast wind of low
density emerges, which leads to the so-called fireworks as the
shell gas is driven out in the form of dense bullets~(McCaughrean
\& Mac Low, 1997). The results presented in this paper corroborate
the recent onset of the formation of a massive star in the ridge
of the RMC, which is in a key transient phase of its emergence
from the natal cloud and the development of the HII region.

\acknowledgments

We are grateful to the referee, Robert Gehrz, for the many helpful
comments and suggestions made. This work was supported by INTAS
grant 4838 and funding from the National Natural Science
Foundation of China through grant 10503006. \\

\begin{table}
\caption{Line fluxes derived from the SOFI/NTT data for the south
and north rims. Note that the 8--6~O(4) and 1--0~S(1) fluxes have
been decomposed by using the CGS4/UKIRT data.} \label{linelist}
$$
\begin{array}{l@{~~~~~}r@{~~~~}r@{~~~~}r@{~~~~~}l@{~~~~~}l@{~~~~~}r@{~~~~}r@{~~~~}r}
\hline \hline \noalign{\smallskip}
\sf{Line}  &   \sf{wavelength} & \sf{flux} &\sf{flux} & & \sf{Line} &  \sf{wavelength} & \sf{flux} &\sf{flux} \\
           & \sf{(\mu m)}      & \sf{South} &\sf{North} & & & \sf{(\mu m)}      & \sf{South} &\sf{North}\\
\noalign{\smallskip} \hline \noalign{\smallskip}
\sf{1-0~S(2)}  &\sf{2.033} &\sf{38.60} & \sf{20.70} & &\sf{4-3~S(6)}      &\sf{2.146} & \sf{<2.00} & \sf{<2.00}\\
\sf{8-6~O(3)}  &\sf{2.041} & \sf{10.30} & \sf{7.04} & &\sf{2-1~S(2)} &\sf{2.154} & \sf{18.20} & \sf{10.82} \\
\sf{3-2~S(5)} &\sf{2.065} & \sf{3.89} & \sf{3.49} & &\sf{9-7~O(2)} &\sf{2.172} & \sf{3.36} & \sf{3.12} \\
\sf{12-9~O(3)} &\sf{2.069} & \sf{<3.00} & \sf{<3.00} & &\sf{3-2~S(3)} &\sf{2.201} & \sf{16.50} & \sf{10.80} \\
\sf{9-7~Q(1)}^\ast &\sf{2.073} & \sf{} & \sf{} & &\sf{4-3~S(5)}^\ast &\sf{2.201} & \sf{} &\sf{} \\
\sf{2-1~S(3)}      &\sf{2.073} & \sf{20.30} & \sf{14.80} & &\sf{8-6~O(5)} &\sf{2.210} & \sf{9.91} & \sf{6.76} \\
\sf{9-7~Q(2)}      &\sf{2.084} & \sf{3.09} & \sf{3.36} & &\sf{1-0~S(0)} &\sf{2.223} &\sf{69.40} &\sf{32.50} \\
\sf{9-7~Q(3)}      &\sf{2.100} & \sf{5.59} & \sf{4.39} & &\sf{2-1~S(1)} &\sf{2.247} &\sf{43.70} &\sf{30.40} \\
\sf{7-5~O(6)}      &\sf{2.108} & \sf{<3.5} & \sf{<3.5} & &\sf{9-7~O(3)} &\sf{2.253} & \sf{10.20} & \sf{6.17} \\
\sf{8-6~O(4)}^\dagger &\sf{2.121} &\sf{10.35} &\sf{5.00} & &\sf{4-3~S(4)} &\sf{2.268} & \sf{4.95} &\sf{<3.00} \\
\sf{1-0~S(1)}^\dagger &\sf{2.121} &\sf{127.65} &\sf{61.60} & &\sf{3-2~S(2)} &\sf{2.286} & \sf{10.00} & \sf{8.16}\\
\sf{3-2~S(4)}      &\sf{2.127} & \sf{10.30} & \sf{5.85} & & & & &\\
\noalign{\smallskip} \hline
\end{array}
$$
Units of $10^{-19}$ W\,m$^{-2}$. The 1$\sigma$ flux measurement
uncertainty is 2\,$\times$\,10$^{-19}$\,W\,m$^{-2}$ and 3.0
$\times$ 10$^{-19}$\,W\,m$^{-2}$ beyond 2.25 $\mu$m. The 12-9~O(3)
flux upper limit
is subject to considerable error due to the location near a strong line.\\
$\ast$ -- Line blended; no decomposition possible. $\dagger$ --
relative line fluxes decomposed from echelle data in approximately
the same regions.
\end{table}

\begin{figure}
\epsscale{0.9} \plotone{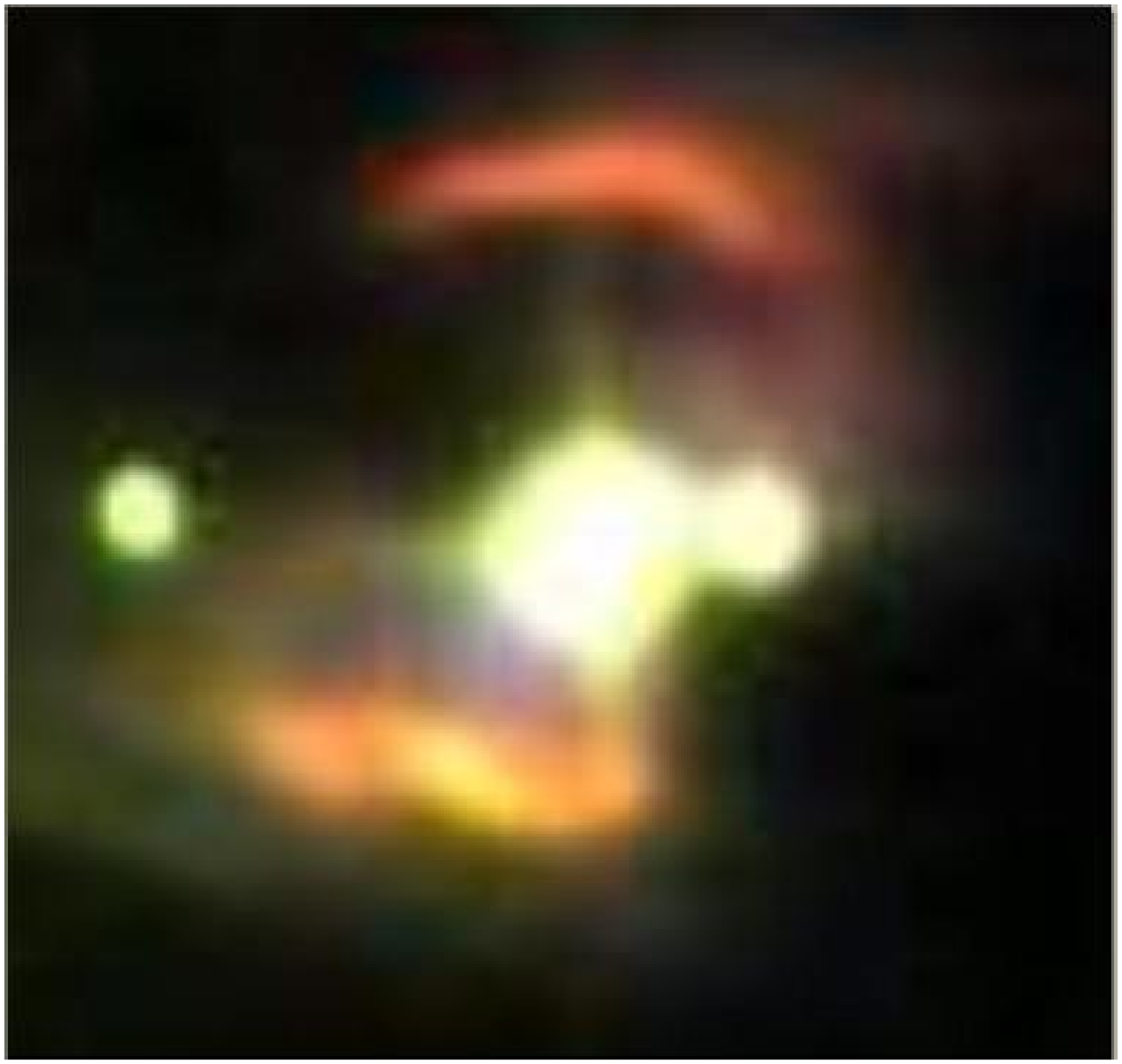} \caption{Close-up view of the
Rosette Eye. North is up and East is to the left. The color
composite with a size of 19\arcsec.0$\times$19\arcsec.0 was
compiled based on the NTT J (blue), H (green) and H$_{2}$ (red)
observations. The image displays both sharp outer features and
inner diffuse emission. The northern rim is considerably redder
and more distant from the central object. Note that the diffuse
fan is stronger in the south with brighter outer edges, suggestive
of a conical shell structure in three dimensions.} \label{closeup}
\end{figure}

\begin{figure}
\epsscale{0.8} \plotone{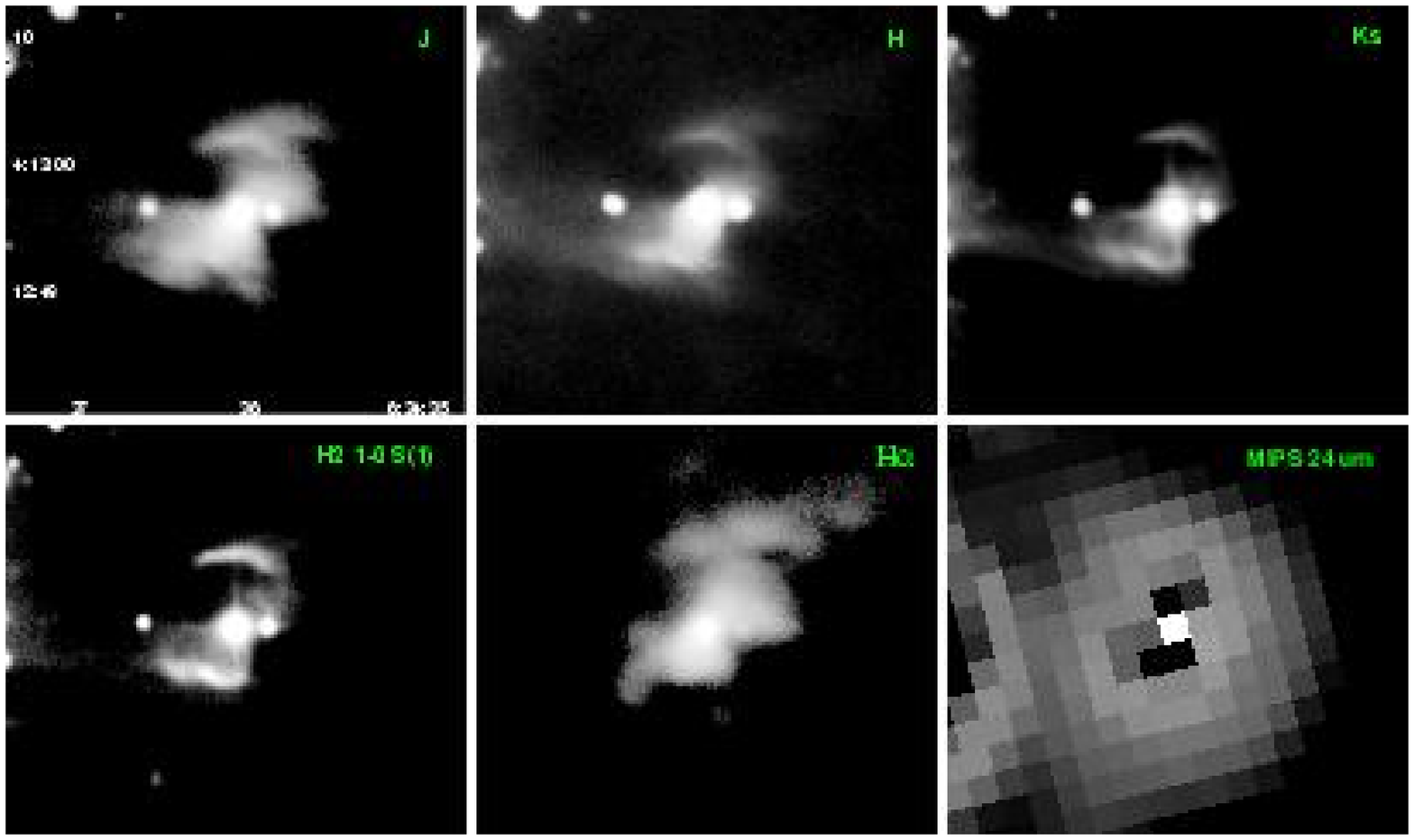} \caption{Greyscale images of
the AFGL\,961~II field in  J (upper-left), H (upper-middle), Ks
(upper-right), 2.12\,$\mu$m (lower-left), H$_{\alpha}$
(lower-middle) \& MIPS 24 $\mu$m (lower-right). North is up and
East is to the left. The morphology of the Eye changes
dramatically between bands. In the J band, it resembles an
hourglass with an eye structure superimposed. In H, Ks and H$_2$,
the Eye is prominent. In H$_{\alpha}$, it indicates predominantly
a fan shaped nebula that extends to the NW and higher extinction
in the opposite direction. The outflow cavity is encompassed by a
dumbbell shaped dusty bubble as disclosed by the MIPS 24 $\mu$m
image.
} \label{quadrants}
\end{figure}

\begin{figure}
\epsscale{1.2} \plottwo{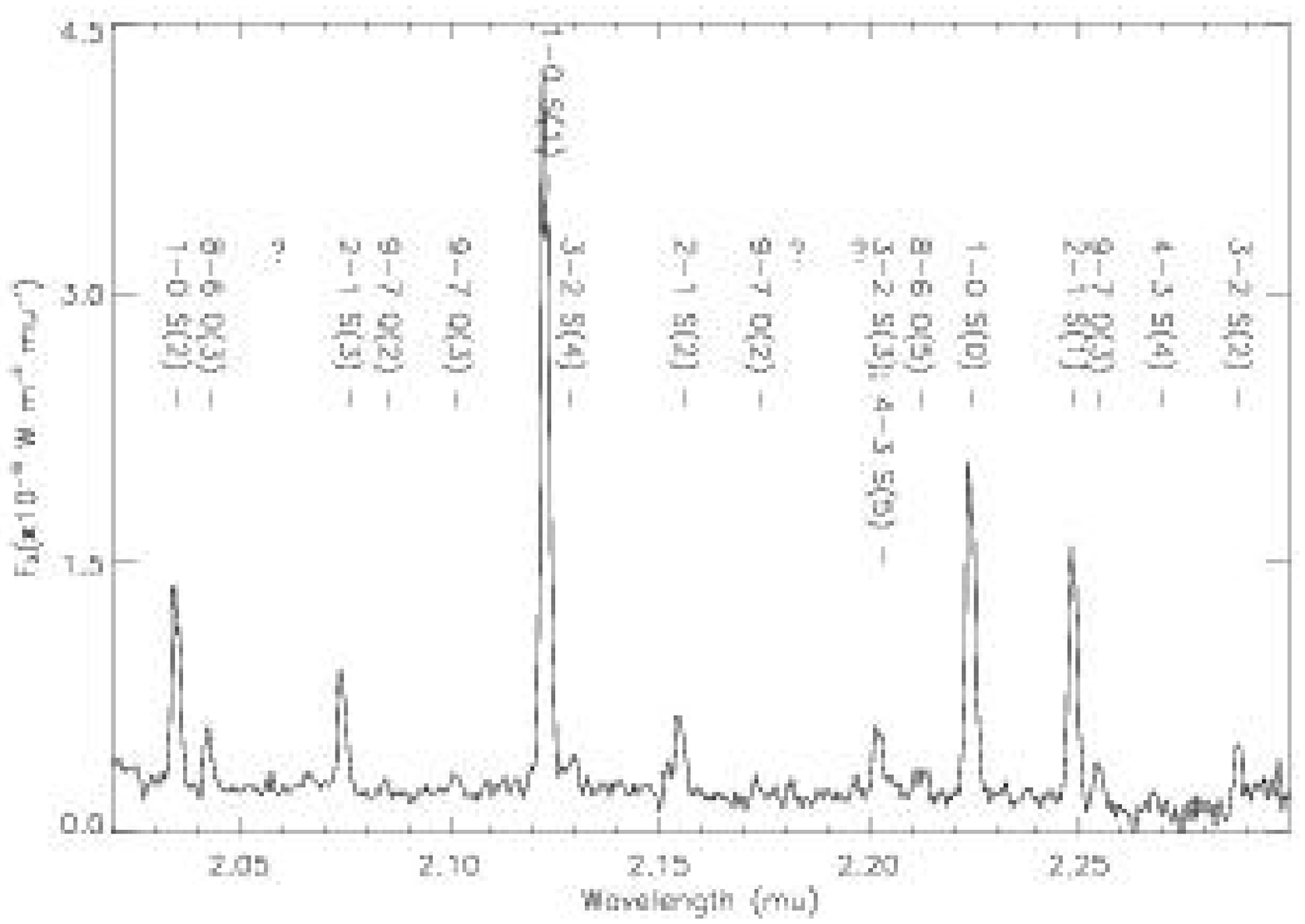}{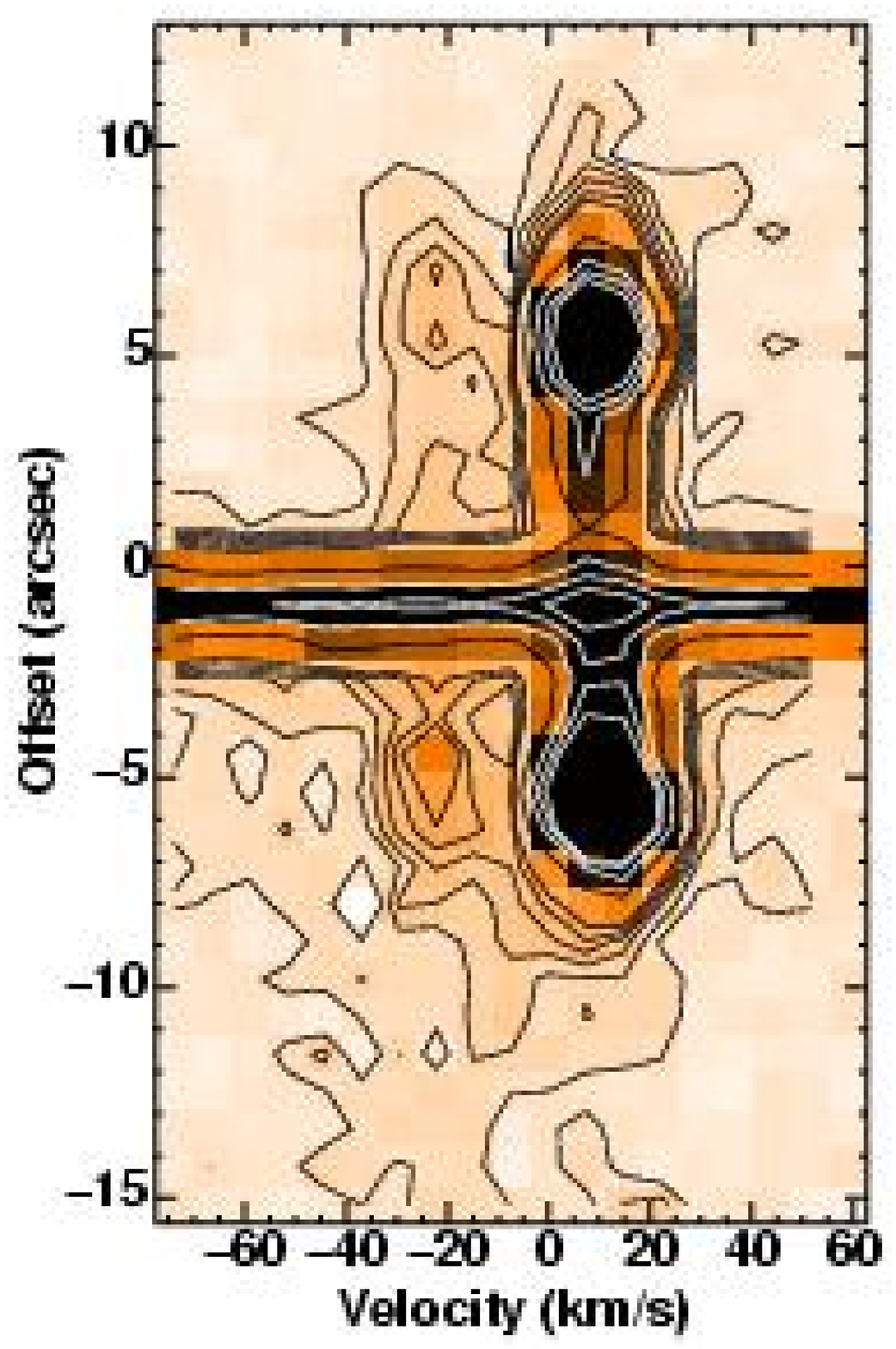} \caption{NTT
Ks-band spectra of the arc structures (upper panel) with the
identified molecular hydrogen lines indicated. It is obvious that
line emission originating from very high vibrational levels of
$H_2$ are present. The highest level present is the 9--7~Q(3) at
2.100\,$\mu$m. It is thus conclusive that there is no indication
of any shock emission at all. Instead, the sharp features are
produced by an exemplary case of fluoresced H$_2$. The
Position-Velocity diagram of the 1--0\,S(1) line emission from the
Ks-band echelle spectroscopy obtained by UKIRT (lower panel)
demonstrates that the material does not possess a measurable
difference in radial velocity, as would be expected if it were
shock-driven. Note that the echelle spectroscopy also uncovers one
further high vibrational line, 3--2\,S(4), that lies very close to
the 1--0\,S(1) line to its red side.} \label{kbandspec}
\end{figure}

\end{document}